\documentstyle[12pt]{article}
\begin{document}
\title{
Path Integral for Relativistic Three-Dimensional
Spinless Aharonov-Bohm-Coulomb System}
\author{De-Hone Lin \thanks{e-mail:d793314@phys.nthu.edu.tw}
\vspace{0.5cm} \\ Department of Physics,
National Tsing Hua University \\
Hsinchu 30043, Taiwan, Republic of China \\}
\maketitle
\setlength{\baselineskip}{1cm} \centerline{\bf Abstract}
{
The path integral for relativistic three-dimensional
spinless Aharonov-Bohm-Coulomb system
is solved, and the energy spectra are extracted from the resulting amplitude.}
\thispagestyle{empty} \newpage
\renewcommand{\thesection}{\arabic{section}}

\tolerance=10000

\section{INTRODUCTION}

{}~~~~~~~In classical mechanics the
effect of an electromagnetic field on a charged
particle is completely described by means of the Lorentz force equation
\begin{equation}
\label{1.1}m\frac{d^2x^\mu }{d\tau ^2}=eF^{\mu \nu }\frac{dx_\nu }{d\tau }
\end{equation}
where $F^{\mu \nu }=\partial ^\mu A^\nu -\partial ^\nu A^\mu $ is the
electromagnetic field tensor and $d\tau =dt(1-v^2)^{1/2}$ is the proper time
interval. The vector
potential $A^\mu $ is merely an auxiliary field. One could 
formulate the particle motion in terms of its derivatives only: the
electric and magnetic fields. 

Surprisingly, it was pointed out in 1959 by AB \cite{1} that, according
to quantum mechanics, the motion of a charged particle can
be influenced by electromagnetic fields in regions from
which the particle is excluded. 
This is called Aharonov-Bohm (AB) effect and 
this effect has been well-confirmed 
experimentally \cite{1.1}. 
In the
past 15 years, AB effect has been studied in the context of anyonic
\cite{2}, cosmic string \cite{3}, and (2+1)-dimensional gravity theories
\cite{4}. Since the anyon, a two dimensional object, carries a magnetic flux
\cite{5}, the dominant interaction between anyons is the AB interaction.
Furthermore, Arovas et al
have pointed out in their seminal paper \cite{6} by calculating the second
virial coefficient 
of the anyons,
that the thermodynamic behavior
of anyon system interpolates between
bosons and fermions. 
An excellent review of past 
theoretical and experimental work 
can be found in
\cite{7}.

Since Feynman propagator is very helpful for analyzing the
scattering and statistical property of anyon system, it is
important to derive the relativistic propagator of the
Aharonov-Bohm-Coulomb (ABC) system. In the past 15 years,
considerable progress has been made in solving path integrals of potential
problems \cite{8,9}. However, only few relativistic problems has been
discussed by PI in \cite{8,10,11,12,12.1,lin,13,13.1}. In this paper, we 
solve the relativistic spinless 3-dimensional ABC system by path integral,
and the energy spectra are extracted from
the resulting amplitude. 

\section{THE RELATIVISTIC PATH INTEGRAL}

~~~~~~~~~Adding
a vector potential ${\bf A}({\bf x})$
to Kleinert's relativistic path integral
for a
particle
in a potential $V({\bf x})$
\cite{8,10},
we find that the expression for the fixed-energy 
amplitude is \cite{lin}
\begin{equation}
\label{3.1}G({\bf x}_b,{\bf x}_a;E)=\frac{i\hbar }{2Mc}\int_0^\infty dL\int
D\rho \Phi \left[ \rho \right] \int D^Dxe^{-A_E/\hbar }
\end{equation}
with the action 
\begin{equation}
\label{3.2}A_E=\int_{\tau _a}^{\tau _b}d\tau \left[ \frac M{2\rho
\left( \tau \right) }{\bf x}^{\prime ^2}\left( \tau \right) -i(e/c)%
{\bf A\cdot x^{\prime }(}\tau {\bf )}-\rho (\tau )\frac{\left(
E-V\right) ^2}{2Mc^2}+\rho \left( \tau \right) \frac{Mc^2}2\right] .
\end{equation}
For the ABC system under consideration, the potential is
\begin{equation}
\label{3.6}V(r)=-e^2/r,
\end{equation}
and the vector potential
\begin{equation}
\label{3.7}A_i=2g\partial _i\theta ,
\end{equation}
where $e$ is the charge and $\theta $ is the azimuthal angle around the
tube:
\begin{equation}
\label{3.8}\theta ({\bf x})=\arctan (x_2/x_1).
\end{equation}
The associated magnetic field lines are confined to an infinitely thin tube
along the $z$-axis:
\begin{equation}
B_3=2g\epsilon _{3jk}\partial _j\partial _k\theta
=2g2\pi \delta ^{(2)}({\bf x}_{\bot }),
\label{3.9}\end{equation}
where ${\bf x}_{\bot }$ is the transverse vector ${\bf x}_{\bot }\equiv
(x_1,x_2).$

Before time-slicing the path integral, we have to regularize it
via a so-called $f$-transformation \cite{8,11}, which exchanges the path
parameter $ \tau$ by a new one $s$:
\begin{equation}
\label{3.3}d\tau =dsf_l({\bf {x}}_n)f_r({\bf {x}}_{n-1}),
\end{equation}
where $f_l({\bf {x}})$ and $f_r({\bf {x}})$ are
invertible functions whose product is positive. The freedom in choosing $%
f_{l,r}$ amounts to an invariance under path-dependent-reparametrizations
of the path parameter $ \tau$
in the fixed-energy amplitude  (\ref{3.1}). By
this transformation, the (D+1)-dimensional relativistic fixed-energy
amplitude for arbitrary time-independent potential turns into \cite{8,11}%
$$
G({\bf x}_b,{\bf x}_a;E)\approx \frac{i\hbar }{2Mc}\int_0^\infty
dS\prod_{n=1}^{N+1}\left[ \int d\rho _n\Phi (\rho _n)\right]
$$
\begin{equation}
\label{3.4}\times \frac{f_l({\bf {x}}_a)f_r({\bf {x}}_b)}{\left( \frac{2\pi
\hbar \epsilon _b^s\rho _bf_l({\bf {x}}_b)f_r({\bf {x}}_a)}M\right) ^{D/2}}%
\prod_{n=1}^N\left[ \int_{-\infty }^\infty \frac{d^Dx_n}{\left( \frac{2\pi
\hbar \epsilon _n^s\rho _nf({\bf {x}}_n)}M\right) ^{D/2}}\right] \exp
\left\{ -\frac 1\hbar A^N\right\}
\end{equation}
with the $s$-sliced action%
$$ \!\!\!\!\!\!\!\!\!\!\!\!\!\!\!\!\!\!\!\!\!\!\!\!\!\!\!
A^N=\sum_{n=1}^{N+1}\left[ \frac{M\left( {\bf {x}}_n-{\bf {x}}_{n-1}\right)
^2}{2\epsilon _n^s\rho _nf_l({\bf {x}}_n)f_r({\bf {x}}_{n-1})}-i\frac{e}{c}{\bf
A}%
_n\cdot ({\bf x}_n-{\bf x}_{n-1})\right.
$$
\begin{equation}
\label{3.5}~~~~~~~~~~~~~~~~~~~\left. -\epsilon _n^s\rho _nf_l({\bf
{x}}_n)f_r({\bf {x}}_{n-1})
\frac{\left( E-V\right) ^2}{2Mc^2}+\epsilon _n^s\rho _nf_l({\bf {x}}_n)f_r({\bf
{x}}_{n-1})\frac{Mc^2}2\right] .
\end{equation}
A family
functions which regulates the ABC system is
\begin{equation}
\label{3.10}f_l({\bf {x}})=f({\bf x})^{1-\lambda },\ \ f_r({\bf {x}})=f({\bf %
x})^\lambda ,
\end{equation}
whose product satisfies $f_l({\bf {x}})f_r({\bf {x}})=f({\bf {x}})=r.$ Thus
arrive at
the amplitude
$$
G({\bf x}_b,{\bf x}_a;E)\approx \frac{i\hbar }{2Mc}\int_0^\infty
dS\prod_{n=1}^{N+1}\left[ \int d\rho _n\Phi (\rho _n)\right]
$$
\begin{equation}
\label{3.11}\times \frac{r_a^{1-\lambda }r_b^\lambda }{\left( \frac{2\pi
\hbar \epsilon _b^s\rho _br_b^{1-\lambda }r_a^\lambda }M\right) ^{3/2}}%
\prod_{n=2}^{N+1}\left[ \int_{-\infty }^\infty \frac{d^3\triangle x_n}{%
\left( \frac{2\pi \hbar \epsilon _n^s\rho _nr_{n-1}}M\right) ^{3/2}}\right]
\exp \left\{ -\frac 1\hbar A^N\right\} ,
\end{equation}
where the action is
$$
A^N=\sum_{n=1}^{N+1}\left[ \frac{M\left( {\bf {x}}_n-{\bf {x}}_{n-1}\right)
^2}{2\epsilon _n^s\rho _nr_n^{1-\lambda }r_{n-1}^\lambda }-i(e/c){\bf A}%
_n\cdot ({\bf x}_n-{\bf x}_{n-1})\right.
$$
\begin{equation}
\label{3.12}\left. -\epsilon _n^s\rho _nr_n\left( r_{n-1}/r_n\right)
^\lambda \frac{(E-V)^2}{2Mc^2}+\epsilon _n^s\rho _nr_n\left(
r_{n-1}/r_n\right) ^\lambda \frac{Mc^2}2\right] .
\end{equation}
For using the Kustaanheimo-Stiefel (KS) transformation (e.g.\cite{8}), we now
incorporate the dummy fourth dimension into the action by replacing ${\bf x\
}$in the kinetic term by the four-vector $\vec x$ and extending the kinetic
action to
\begin{equation}
\label{3.12.1}A_{{\rm kin}}^N=\sum_{n=1}^{N+1}\frac M2\frac{\left( \vec
x_n-\vec x_{n-1}\right) }{\epsilon _n^s\rho _nr_n^{1-\lambda
}r_{n-1}^\lambda }.
\end{equation}
This is achieved by inserting the following trivial identity
\begin{equation}
\label{3.12.2}\prod_{n=1}^{N+1}\left[ \int \frac{d(\triangle x^4)_n}{\left(
2\pi \hbar \epsilon _n^s\rho _nr_n^{1-\lambda }r_{n-1}^\lambda /M\right)
^{1/2}}\right] \exp \left\{ -\frac 1\hbar \sum_{n=1}^{N+1}\frac M2\frac{%
(\triangle x_n^4)^2}{\epsilon _n^s\rho _nr_n^{1-\lambda }r_{n-1}^\lambda }%
\right\} =1.
\end{equation}
Hence the fixed-energy amplitude of the ABC system in three dimensions
can be rewritten as the four-dimensional path integral%
$$
G({\bf x}_b,{\bf x}_a;E)\approx \frac{i\hbar }{2Mc}\int_0^\infty
dS\prod_{n=1}^{N+1}\left[ \int d\rho _n\Phi (\rho _n)\right]
$$
\begin{equation}
\label{3.12.3}\times \int dx_a^4\frac{r_a^{1-\lambda }r_b^\lambda }{\left(
\frac{2\pi \hbar \epsilon _b^s\rho _br_b^{1-\lambda }r_a^\lambda }M\right) ^2%
}\prod_{n=2}^{N+1}\left[ \int_{-\infty }^\infty \frac{d^4\triangle x_n}{%
\left( \frac{2\pi \hbar \epsilon _n^s\rho _nr_{n-1}}M\right) ^2}\right] \exp
\left\{ -\frac 1\hbar A^N\right\} ,
\end{equation}
where $A^N$ is the action Eq. (\ref{3.12}) in which the three-vectors ${\bf x%
}_n$ are replaced by the four-vectors $\vec x_n$. With the help of the
following approximation%
$$
\frac{r_a^{1-\lambda }r_b^\lambda }{\left( \frac{2\pi \hbar \epsilon
_b^s\rho _br_b^{1-\lambda }r_a^\lambda }M\right) ^2}\prod_{n=2}^{N+1}\left[
\int_{-\infty }^\infty \frac{d^4\triangle x_n}{\left( \frac{2\pi \hbar
\epsilon _n^s\rho _nr_{n-1}}M\right) ^2}\right]
$$
\begin{equation}
\label{3.12.4}\approx \frac 1{r_a}\frac 1{\left( \frac{2\pi \hbar \epsilon
_b^s\rho _b}M\right) ^2}\prod_{n=2}^{N+1}\left[ \int_{-\infty }^\infty \frac{%
d^4\triangle x_n}{\left( \frac{2\pi \hbar \epsilon _n^s\rho _nr_n}M\right) ^2%
}\right] \exp \left\{ 3\lambda \sum_{n=1}^{N+1}\log \frac{r_n}{r_{n-1}}%
\right\} ,
\end{equation}
where the equality $(r_b/r_a)^{3\lambda
-2}=\prod_1^{N+1}(r_n/r_{n-1})^{3\lambda -2}$ has been used, we arrive%
$$
G({\bf x}_b,{\bf x}_a;E)\approx
 \frac{i\hbar }{2Mc}\int_0^\infty
dS\prod_{n=1}^{N+1}\left[ \int d\rho _n\Phi (\rho _n)\right]
$$
\begin{equation}
\label{3.12.5}\times \int dx_a^4\frac 1{r_a}\frac 1{\left( \frac{2\pi \hbar
\epsilon _b^s\rho _b}M\right) ^2}\prod_{n=2}^{N+1}\left[ \int_{-\infty
}^\infty \frac{d^4\triangle x_n}{\left( \frac{2\pi \hbar \epsilon _n^s\rho
_nr_n}M\right) ^2}\right] \exp \left\{ -\frac 1\hbar \sum_{n=1}^{N+1}\left[
A^N-3\lambda \hbar \log \frac{r_n}{r_{n-1}}\right] \right\} .
\end{equation}
Since the path integral represents the general relativistic resolvent
operator, all results must be independent of the splitting parameter $%
\lambda $ after going to the continuum limit. Choosing a splitting parameter
$\lambda =0,$ we obtain the continuum limit of the action
\begin{equation}
\label{3.13}A_E\left[ x,x^{\prime }\right] =\int_0^Sds\left[ \frac{%
Mx^{\prime 2}}{2\rho r}-i(e/c){\bf A}\cdot {\bf x}^{\prime }-\rho r\frac{%
(E-V)^2}{2Mc^2}+\rho r\frac{Mc^2}2\right]
\end{equation}
We now solve the ABC system by introducing the KS transformation (e.g.\cite{8})
\begin{equation}
\label{3.14}d\vec x=2A(\vec u)d\vec u.
\end{equation}
The arrow on top of $x$ indicates that $x$ has become a four-vector. For
symmetry reasons, we choose the $4\times 4$ matrix $A(\vec u)$ as
\begin{equation}
\label{3.14.1}A(\vec u)=\left(
\begin{array}{cccc}
u^3 & u^4 & u^1 & u^2 \\
u^4 & -u^3 & -u^2 & u^1 \\
u^1 & u^2 & -u^3 & -u^4 \\
u^2 & -u^1 & u^4 & -u^3
\end{array}
\right) .
\end{equation}
The transformation of coordinate difference is
\begin{equation}
\label{3.17}(\triangle {\bf x}_n^i)^2=4{\bf \bar u}_n^2(\triangle {\bf u}%
_n^i)^2,
\end{equation}
where ${\bf \bar u}_n$ $\equiv ({\bf u}_n+{\bf u}_{n-1})/2$. In the
continuum limit, this amounts to
\begin{equation}
\label{3.18}d^4x_n=16{\bf u}_n^2d^4u_n=16r_n^2d^4u_n,
\end{equation}
\begin{equation}
\label{3.18.1}\vec x^{\prime 2}=4\vec u^2\vec u^{\prime 2}=4r\vec u^{\prime
2}.
\end{equation}
By employing the basis tetrad notation $e_{\;\;\mu }^i(\vec u),$ Eq. (\ref
{3.14}) has the form $dx^i=e_{\;\;\mu }^i(\vec u)$ $du^\mu $, this is given
by
\begin{equation}
\label{3.19}e_{\;\;\mu }^i(\vec u)\ =\frac{\partial x^i}{\partial u^\mu }%
(\vec u)=2A_{\;\ \mu }^i(\vec u)\;,\quad i=1,2,3,4.
\end{equation}
Under the KS transformation, the magnetic interaction turns into%
$$
{\bf A}_n\cdot ({\bf x}_n-{\bf x}_{n-1})=-2g\frac{x_n^2\triangle
x_n^1-x_n^1\triangle x_n^2}{r_n^2}
$$
\begin{equation}
\label{3.20}=-2g\left[ \frac{u_n^1\triangle u_n^2-u_n^2\triangle u_n^1}{%
\left( u_n^1\right) ^2+\left( u_n^1\right) ^2}+\frac{u_n^4\triangle
u_n^3-u_n^3\triangle u_n^4}{\left( u_n^3\right) ^2+\left( u_n^4\right) ^2}%
\right] .
\end{equation}
We obtain a path integral equivalent to Eq. (\ref{3.12.3})
\begin{equation}
\label{3.21}G({\bf x}_b,{\bf x}_a;E)= \frac{i\hbar }{2Mc}\int_0^\infty
dS\ e^{SEe^2/\hbar Mc^2}G(\vec u_b,\vec u_a;S),
\end{equation}
where $G(\vec u_b,\vec u_a;S)$ denotes the s-sliced amplitude
\begin{equation}
\label{3.21.5}\prod_{n=1}^{N+1}\left[ \int d\rho _n\Phi (\rho _n)\right]
\frac 1{16}\int \frac{dx_a^4}{r_a}\frac 1{\left( \frac{2\pi \hbar \epsilon
_b^s\rho _b}m\right) ^2}\prod_{n=1}^N\left[ \int_{-\infty }^\infty \frac{%
d^4u_n}{\left( \frac{2\pi \hbar \epsilon _n^s\rho _n}m\right) ^2}\right]
\exp \left\{ -\frac 1\hbar A^N\right\}
\end{equation}
with the action
\begin{equation}
\label{3.22}A^N=\sum_{n=1}^{N+1}\left\{ \frac{m(\triangle \vec u_n)^2}{%
2\epsilon _n^s\rho _n}-i(e/c)(\vec A_n\cdot \triangle \vec u_n)+\epsilon
_n^s\rho _n\frac{m\omega ^2\vec u_n^2}2-\epsilon _n^s\rho _n\frac{\hbar
^24\alpha ^2}{2m\vec u_n^2}\right\} .
\end{equation}
Here
\begin{equation}
\label{3.23}
m=4M,\quad \omega ^2=\frac{M^2c^4-E^2}{4M^2c^2},
\end{equation}
and
\begin{equation}
\label{3.23.1}\vec A_n\cdot \triangle \vec u_n=-2g\left[ \frac{%
u_n^1\triangle u_n^2-u_n^2\triangle u_n^1}{\left( u_n^1\right) ^2+\left(
u_n^1\right) ^2}+\frac{u_n^4\triangle u_n^3-u_n^3\triangle u_n^4}{\left(
u_n^3\right) ^2+\left( u_n^4\right) ^2}\right] .
\end{equation}
We now choose the gauge $\rho (s)=1$ in Eq. (\ref{3.21.5}). This leads to the
Duru-Kleinert transformed action
\begin{equation}
\label{3.34}A=\int_0^Sds\left[ \frac{m\vec u^{\prime 2}}2-2i(e/c)(\vec
A\cdot \vec u^{\prime })+\frac{m\omega ^2\vec u^2}2-\frac{4\hbar ^2\alpha ^2
}{2m\vec u^2}\right] .
\end{equation}
It describes a particle, forgetting the magnetic interaction term
for a while, of mass $%
m=4M$ moving as a function of the ``pseudotime'' $s$ in a 4-dimensional
harmonic oscillator potential of frequency
\begin{equation}
\label{3.35}\omega ^2=\frac{M^2c^4-E^2}{4M^2c^2}.
\end{equation}
The oscillator possesses an additional attractive potential -$4\hbar
^2\alpha ^2/2m\vec u^2$ which is conveniently parametrized in the form of a
centrifugal barrier
\begin{equation}
\label{3.36}V_{{\rm extra}}=\hbar ^2\frac{l_{{\rm extra}}^2}{2m\vec u^2},
\end{equation}
whose squared angular momentum has the negative value $l_{{\rm extra}%
}^2\equiv -4\alpha ^2,$ where $\alpha $ denotes the fine-structure constant $%
\alpha \equiv e^2/\hbar c\approx 1/137$.

There are no $\lambda $-slicing corrections. This is ensured by the affine
connection of KS transformation satisfying
\begin{equation}
\label{3.24}\Gamma _\mu ^{\;\;\;\mu \lambda }=g^{\mu \nu }e_i^{\;\;\lambda
}\partial _\mu e_{\;\;\nu }^i=0
\end{equation}
and the transverse gauge $\partial _\mu A^\mu =0$ \cite{8,10}. We now analyze
the effect come from the magnetic interaction. Note that the system
separable like $R^4\rightarrow R^2\times R^2$ if the centrifugal term
is not considered for a while. Therefore the path integral in $u$ space become
two independent two-dimensional AB plus harmonic oscillator. This makes the
path integral calculation of $G(\vec u_b,\vec u_a;S)$ extremely simple. For
each two-dimensional system, the derivatives in front of $\varphi $ in Eq. (%
\ref{3.9}) commute everywhere, except at the origin where Stokes' theorem
yields
\begin{equation}
\label{3.25}\int d^2u(\partial _1\partial _2-\partial _2\partial _1)\varphi
=\oint d\varphi =2\pi
\end{equation}
The magnetic flux through the tube is defined by the integral
\begin{equation}
\label{3.26}\Phi =\int d^2uB_3.
\end{equation}
A comparison with Eq. (\ref{3.9}) shows that the coupling constant in Eq. (%
\ref{3.7}) is related to the magnetic flux by
\begin{equation}
\label{3.27}g=\frac \Phi {4\pi }.
\end{equation}
When inserting $A_i=2g\partial _i\varphi $ into Eq. (\ref{3.34}), the
interaction takes the form
\begin{equation}
\label{3.28}A_{{\rm mag}}=-\hbar \beta _0\int_0^Sds\varphi ^{\prime },
\end{equation}
where $\beta _{0}$ is the dimensionless number
\begin{equation}
\label{3.29}\beta _{0}\equiv -\frac{2eg}{\hbar c}.
\end{equation}
The minus sign is a matter of convention. Since the particle orbits are
present at all times, their worldlines in spacetime can be considered as
being closed at infinity, and the integral
\begin{equation}
\label{3.30}n=\frac 1{2\pi }\int_0^Sds\varphi ^{\prime }
\end{equation}
is the topological invariant with integer values of the winding number $n$.
The magnetic interaction is therefore a purely topological one, its value
being
\begin{equation}
\label{3.31}A{\rm _{mag}}=-\hbar \beta _02\pi n.
\end{equation}
After adding this to the action of Eq. (\ref{3.34}) in the radial
decomposition of the relativistic path integral \cite{8,11}, we rewrite the
sum over the azimuthal quantum numbers $m$ via Poisson's summation formula,
and obtain%
$$
G({\bf u}_b,{\bf u}_a;S)=\int_{-\infty }^\infty d\beta \frac 1{\sqrt{u_bu_a}%
}G(u_b,u_a;S)_\beta
$$
\begin{equation}
\label{3.32}\times \sum_{n=-\infty }^\infty \frac 1{2\pi }e^{i(\beta -\beta
_0)(\varphi _b+2n\pi -\varphi _a)}.
\end{equation}
Since the winding number $n$ is often not easy to measure experimentally,
let us extract observable consequences which are independent of $n.$ The sum
over all $n$ forces $\beta $ to be equal to $\beta _0$ modula an arbitrary
integer number. the result, for each $R^2$, is
\begin{equation}
\label{3.33}G({\bf u}_b,{\bf u}_a;S)=\sum_{k=-\infty }^\infty \frac 1{\sqrt{%
u_bu_a}}G(u_b,u_a;S)_{k+\beta _0}\frac 1{2\pi }e^{ik(\varphi _b-\varphi _a)}.
\end{equation}
Therefore we obtain the fixed-energy amplitude%
$$
G({\bf x}_b,{\bf x}_a;E)= \frac{i\hbar }{2Mc}\int_0^\infty dS\
e^{SEe^2/\hbar Mc^2}\frac 1{16}\int \frac{dx_a^4}{r_a}\left( \frac{m\omega }{%
\hbar \sinh \omega s}\right) ^2
$$
$$
\times \sum_{k_1=-\infty }^\infty \sum_{k_2=-\infty }^\infty e^{ik_1(\varphi
_{1,b}-\varphi _{1,a})}e^{ik_1(\varphi _{2,b}-\varphi _{2,a})}
$$
$$
\times \exp \left\{ -\frac{m\omega }{2\hbar }\left( \sigma _{1,b}^2+\sigma
_{1,a}^2+\sigma _{2,b}^2+\sigma _{2,a}^2\right) \coth \omega s\right\}
$$
\begin{equation}
\label{3.35.1}\times I_{\mid k_1+\beta _0\mid }\left( \frac m\hbar \frac{%
\omega \sigma _{1,b}\sigma _{1,a}}{\sinh \omega s}\right) I_{\mid k_2+\beta
_0\mid }\left( \frac m\hbar \frac{\omega \sigma _{2,b}\sigma _{2,a}}{\sinh
\omega s}\right) ,
\end{equation}
where $(\sigma _1,\varphi _1)$ and $(\sigma _2,\varphi _2)$ are defined by
\begin{equation}
\label{3.25.2}\left.
\begin{array}{c}
u^1=\sigma _1\sin \varphi _1 \\
u^2=\sigma _1\cos \varphi _1 \\
u^3=\sigma _2\cos \varphi _2 \\
u^4=\sigma _2\sin \varphi _2
\end{array}
\right\} .
\end{equation}
In order to perform the $x_a^4$-integration we express $(\sigma _1,\varphi
_1,\sigma _2,\varphi _2)$ in terms of three-dimensional spherical coordinate
with an auxiliary angle $\gamma $:
\begin{equation}
\label{3.15.3}\left.
\begin{array}{c}
u^1=
\sqrt{r}\cos (\theta /2)\cos \left[ (\varphi +\gamma )/2\right]  \\ u^2=
\sqrt{r}\cos (\theta /2)\sin \left[ (\varphi +\gamma )/2\right]  \\ u^3=
\sqrt{r}\sin (\theta /2)\cos \left[ (\varphi -\gamma )/2\right]  \\ u^4=
\sqrt{r}\sin (\theta /2)\sin \left[ (\varphi -\gamma )/2\right]
\end{array}
\right\} \qquad \left(
\begin{array}{c}
0\leq \theta \leq \pi  \\
0\leq \varphi \leq 2\pi  \\
0\leq \gamma \leq 4\pi
\end{array}
\right)
\end{equation}
and identify

\begin{equation}
\label{3.25.4}\left.
\begin{array}{l}
\sigma _1=
\sqrt{r}\cos (\theta /2) \\ \varphi_1=(\varphi +\gamma +\pi )/2 \\
\sigma _2=
\sqrt{r}\sin (\theta /2) \\ \varphi _2=(\varphi -\gamma )/2
\end{array}
\right\} .
\end{equation}
Then one can change the $x_a^4$-integration into the $\gamma _a$-integration
whose result is easily represented as the Kronecker delta $\delta _{k_1,k_2}$%
. Hence, one can carry out $k_2$-summation and finally becmes%
$$
G({\bf x}_b,{\bf x}_a;E)= \frac{i\hbar }{2Mc}\frac{M^2\omega }{\pi
\hbar^2 }\sum_{k=-\infty }^\infty e^{ik(\varphi _b-\varphi _a)}
$$
$$
\times \int_0^\infty dye^{2(Ee^2/2\omega \hbar Mc^2)y}\frac 1{\sinh
{}^2y}e^{-\frac{m\omega }{2\hbar }(r_b+r_a)\coth y}
$$
\begin{equation}
\label{3.25.5}\times I_{\mid k+\beta _0\mid }\left( \frac{m\omega \sqrt{%
r_br_a}}{\hbar \sinh y}\cos \theta _b/2\cos \theta _a/2\right) I_{\mid
k+\beta _0\mid }\left( \frac{m\omega \sqrt{r_br_a}}{\hbar \sinh y}\sin
\theta _b/2\sin \theta _a/2\right) ,
\end{equation}
where we have defined the new variabe $y=\omega s$.
We make now use of the addition theorem Ref. \cite{14}, Vol. II, p.99:
\begin{equation}
\frac z2J_\nu \left( z\sin \alpha \sin \beta \right) J_\mu \left( z\cos
\alpha \cos \beta \right)
\end{equation}
$$
=\left( \sin \alpha \sin \beta \right) ^\nu \left( \cos \alpha \cos \beta
\right) ^\mu \sum_{l=0}^\infty (-1)^l\left( \mu +\nu +2l+1\right)
$$
$$
\times \frac{\Gamma \left( \mu +\nu +l+1\right) \Gamma \left( \nu
+l+1\right) }{l!\Gamma \left( \mu +l+1\right) \Gamma ^2\left( \nu +1\right) }%
J_{\mu +\nu +l+1}\left( z\right)
$$
$$
\times \;_2F_1\left( -l,\mu +\nu +l+1,\nu +1;\sin {}^2\alpha \right)
$$
\begin{equation}
\label{3.25.6}\times \;_2F_1\left( -l,\mu +\nu +l+1,\nu +1;\sin {}^2\beta
\right) ,
\end{equation}
and the relation between hypergeometrc function and Jacobi-polymomials
\begin{equation}
\label{3.15.7}P_l^{(\alpha ,\beta )}(z)=\frac{\Gamma \left( \alpha
+l+1\right) }{\Gamma \left( \alpha +1\right) l!}\;_2F_1\left( \alpha +\beta
+l+1,-l;1+\alpha ;\frac{1-z}2\right) .
\end{equation}
We arrive
$$
G({\bf x}_b,{\bf x}_a;E)= \frac{i\hbar }{2Mc}\frac M{2\pi \hbar \sqrt{r_br_a}%
}\sum_{k=-\infty }^\infty \sum_{n=0}^\infty e^{ik(\varphi _b-\varphi _a)}
$$
$$
\times \left( \cos \theta _b/2\cos \theta _a/2\right) ^{\mid k+\beta _0\mid
}\left( \sin \theta _b/2\sin \theta _a/2\right) ^{\mid k+\beta _0\mid }
$$
$$
\times \frac{n!\Gamma \left( n+2\mid k+\beta _0\mid +1\right) \left(
2n+2\mid k+\beta _0\mid +1\right) }{\Gamma ^2\left( n+2\mid k+\beta _0\mid
+1\right) }
$$
$$
\times \left\{ \vbox to 24pt{}
 \int_0^\infty dye^{2(Ee^2/2\omega \hbar Mc^2)y}\frac 1{\sinh
{}y}e^{-\frac{m\omega }{2\hbar }(r_b+r_a)\coth y}I_{2n+2\mid k+\beta
_0\mid +1}\left( \frac{m\omega \sqrt{r_br_a}}{\hbar \sinh y}\right)
\vbox to 24pt{} \right\}
$$
\begin{equation}
\label{3.15.8}\times P_n^{(\mid k+\beta _0\mid ,\mid k+\beta _0\mid )}\left(
\cos \theta _b\right) P_n^{(\mid k+\beta _0\mid ,\mid k+\beta _0\mid )}\left(
\cos \theta _a\right) .
\end{equation}
At this place, the additional centrifugal barrier (\ref{3.36}) is
incorporated via the replacement \cite{8}
\begin{equation}
\label{3.15.9}\left( 2n+2\mid k+\beta _0\mid +1\right) \rightarrow \sqrt{%
\left( 2n+2\mid k+\beta _0\mid +1\right) ^2-4\alpha ^2}.
\end{equation}

This integral can be calculated by employing the formula%
$$
\int_0^\infty dy\frac{e^{2\nu y}}{\sinh y}\exp \left[ -\frac t2\left( \zeta
_a+\zeta _b\right) \coth y\right] I_\mu \left( \frac{t\sqrt{\zeta _b\zeta _a}
}{\sinh y}\right)
$$
\begin{equation}
\label{3.41}=\frac{\Gamma \left( \left( 1+\mu \right) /2-\nu \right) }{t
\sqrt{\zeta _b\zeta _a}\Gamma \left( \mu +1\right) }W_{\nu ,\mu /2}\left(
t\zeta _b\right) M_{\nu ,\mu /2}\left( t\zeta _b\right) ,
\end{equation}
with the range of validity%
$$
\begin{array}{l}
\zeta _b>\zeta _a>0, \\
{Re}[(1+\mu )/2-\nu ]>0, \\ {Re}(t)>0,\mid \arg t\mid <\pi ,
\end{array}
$$
where $M_{\mu ,\nu }$ and $W_{\mu ,\nu }$ are the Whittaker functions, we
complete the integration of Eq. (\ref{3.15.8}), and find the amplitude for $%
u_b>u_a$ in the closed form,%
$$
G({\bf x}_b,{\bf x}_a;E)= \frac {i\hbar} {2Mc}\frac{M c}{4\pi r_br_a
\sqrt{M^2c^4-E^2}}\sum_{k=-\infty }^\infty \sum_{n=0}^\infty e^{ik(\varphi
_b-\varphi _a)}
$$
$$
\times \left( \cos \theta _b/2\cos \theta _a/2\right) ^{\mid k+\beta _0\mid
}\left( \sin \theta _b/2\sin \theta _a/2\right) ^{\mid k+\beta _0\mid }
$$
$$
\times \frac{n!\Gamma \left( n+2\mid k+\beta _0\mid +1\right) \left(
2n+2\mid k+\beta _0\mid +1\right) }{\Gamma ^2\left( n+2\mid k+\beta _0\mid
+1\right) }
$$
$$
\times \frac{\Gamma \left( \frac 12+\frac 12\sqrt{\left( 2n+2\mid k+\beta
_0\mid +1\right) ^2-4\alpha ^2}-\frac{E\alpha }{\sqrt{M^2c^4-E^2}}\right) }{%
\Gamma \left( \sqrt{\left( 2n+2\mid k+\beta _0\mid +1\right) ^2-4\alpha ^2}%
+1\right) }
$$
$$
\times W_{E\alpha /\sqrt{M^2c^4-E^2},\sqrt{(2n+2\mid k+\beta _0\mid
+1)^2-4\alpha ^2}/2}\left( \frac 2{\hbar c}\sqrt{M^2c^4-E^2}r_b\right)
$$
$$
\times M_{E\alpha /\sqrt{M^2c^4-E^2},\sqrt{(2n+2\mid k+\beta _0\mid
+1)^2-4\alpha ^2}/2}\left( \frac 2{\hbar c}\sqrt{M^2c^4-E^2}r_a\right)
$$
\begin{equation}
\label{3.15.8}\times P_n^{(\mid k+\beta _0\mid ,\mid k+\beta _0\mid) }\left(
\cos \theta _b\right) P_n^{(\mid k+\beta _0\mid ,\mid k+\beta _0\mid) }\left(
\cos \theta _a\right) .
\end{equation}
The energy spectra can be extracted from the poles. They are determined by
\begin{equation}
\label{3.43}\frac 12+\frac 12\sqrt{\left( 2n+2\mid k+\beta _0\mid +1\right)
^2-4\alpha ^2}-\frac{E\alpha }{\sqrt{M^2c^4-E^2}}=-n_r,\;n_r=0,1,2,\cdots .
\end{equation}
Expanding this equation into a power of $\alpha ,$ we get
$$
E_{n_r,n,k}=\pm Mc^2 \left\{ \vbox to 24pt{}
 1-\frac 12\left( \frac \alpha {n_r+n+\mid k+\beta
_0\mid +1}\right) ^2-\frac{\alpha ^4}{\left( n_r+n+\mid k+\beta _0\mid
+1\right) ^3}\right.
$$
\begin{equation}
\label{3.44}\left. \times \left[ \frac 1{2n+2\mid k+\beta _0\mid +1}-\frac
3{8\left( n_r+n+\mid k+\beta _0\mid +1\right) }\right] +O\left( \alpha
^6\right) \vbox to 24pt{} \right\}
  ,\quad n_r,n=0,1,2,3,\cdots
\end{equation}
In the non-relativistic limit, the spectra is in agreement with the result
in Ref. \cite{17,15,16}.

The alert reader will have noted the similarity of the
techniques used in this paper
to those leading to
the solution of the path integral of the dionium atom \cite{18}.
\\
\centerline{ACKNOWLEDGMENTS}
\\
The author is grateful to Professor  
H. Kleinert who 
critically read the entire manuscript and made corrections.

\end{document}